\documentclass[aps,pra,showkeys,twocolumn,superscriptaddress]{revtex4-2}
\usepackage{array}
\usepackage{booktabs}
\usepackage{tabu}
\usepackage{dcolumn}
\usepackage{amsmath}
\usepackage{amsfonts}
\usepackage{float}
\usepackage{amssymb}
\usepackage{graphicx,color}
\usepackage[colorlinks={true}]{hyperref}
\hypersetup{colorlinks=true,linkcolor=red,citecolor=blue,urlcolor=blue}
\usepackage{graphicx}
\usepackage{subfigure}
\usepackage{graphicx}
\usepackage{dcolumn}
\usepackage{bm}
\usepackage{pstricks}
\usepackage{braket}
\usepackage{orcidlink}

\def\be{\begin{equation}}
  \def\ee{\end{equation}}
\def\bea{\begin{eqnarray}}
\def\eea{\end{eqnarray}}
\def\f{\frac}
\def\n{\nonumber}
\def\l{\label}
\def\p{\phi}
\def\o{\over}
\def\R{\rho}
\def\pa{\partial}
\def\om{\omega}
\def\na{\nabla}
\def\P{\Phi}
\begin{document}

\title{Practical Scheme for Realization of a Quantum Battery}

\author{Maryam Hadipour \orcidlink{0000-0002-6573-9960}}
\affiliation{Faculty of Physics, Urmia University of Technology, Urmia, Iran}
\author{Soroush Haseli \orcidlink{0000-0003-1031-4815}}\email{soroush.haseli@uut.ac.ir}
\affiliation{Faculty of Physics, Urmia University of Technology, Urmia, Iran}
\affiliation{School of Physics, Institute for Research in Fundamental Sciences (IPM),  P.O. Box 19395-5531, Tehran, Iran}
\author{Dong Wang \orcidlink{0000-0002-0545-6205}}\email{dwang@ahu.edu.cn}
\affiliation{School of Physics and Optoelectronics Engineering, Anhui University, Hefei, 230601, People’s Republic of China}
\author{Saeed Haddadi \orcidlink{0000-0002-1596-0763}}\email{haddadi@semnan.ac.ir}
\affiliation{Faculty of Physics, Semnan University, P.O. Box 35195-363, Semnan, Iran}

\date{\today}
\def\be{\begin{equation}}
  \def\ee{\end{equation}}
\def\bea{\begin{eqnarray}}
\def\eea{\end{eqnarray}}
\def\f{\frac}
\def\n{\nonumber}
\def\l{\label}
\def\p{\phi}
\def\o{\over}
\def\R{\rho}
\def\pa{\partial}
\def\om{\omega}
\def\na{\nabla}
\def\P{$\Phi$}

\begin{abstract}
We propose a practical scheme for a quantum battery consisting of an atom-cavity interacting system under a structured reservoir in the non-Markovian regime. We investigate a multi-parameter regime for the cavity-reservoir coupling and reveal how these parameters affect the performance of the quantum battery. Our proposed scheme is simple and may be achievable for practical realization and implementation.
\end{abstract}


\maketitle

\section{INTRODUCTION}	
Traditional batteries that are still in use, such as lithium-ion, alkaline, and lead-acid batteries, operate based on electrochemical reactions that involve the motion of ions between two electrodes through an electrolyte. The performance of these batteries strongly depends on various factors such as the electrolyte composition, the materials used in electrodes, and the overall design. Quantum batteries (QBs) \cite{qb1}, on the other hand, are a novel concept that probes the potential of quantum mechanics to enhance energy storage. These batteries, at least in theory, can use quantum superposition and entanglement to store and recover energy more efficiently than traditional batteries \cite{qb2,qb3,new1,new2,new3}. However, QBs are still in the early stages of development and face notable challenges in terms of stability, scalability, and practical implementation \cite{qb4,qb5,qb6}.

A typical model of the QB comprises two components, the battery charger and the battery holder. The latter, designed to prevent energy loss, is essentially isolated from the surrounding environment and is treated as a dissipation-less subsystem. To obtain energy, the battery holder must be coupled to the battery charger. After a limited charging period, the battery holder detaches from the battery charger, allowing energy storage and ultimately extraction of the required energy \cite{qb7}. Interestingly, this basic bipartite model and other models of QB, along with its theoretical explorations in various forms, have been the subject of intensive QB research in recent years \cite{qb8,qb9,qb10,qb11,qb12,qb13,qb14,qb15,qb16,qb17,qb18,qb19,qbnew1,qbnew2,qb20,qb21,qb22,qb23}.

QBs illustrate an area of research at the intersection of quantum physics and energy storage, however, as mentioned above, their development has some potential challenges and limitations. For example, QBs rely on preserving quantum coherence \cite{Coherence0,Coherence1,Coherence2,Coherence3,Coherence4}, which is the ability of quantum systems to exist in a superposition of states. Practically speaking, preserving quantum coherence for a sufficient period of time can be difficult due to some sources of decoherence \cite{decoh1,decoh2,decoh3,decoh4,decoh5,decoh6,decoh7,decoh8,decoh9,decoh10,decoh11,decoh12,decoh13}. In other words, a significant challenge is constructing a QB that can resist external effects and maintain quantum coherence in a real-world environment. Therefore, addressing environmental effects on the performance of QBs is necessary for their practical implementation and development.

In this paper, the presence of a mediated cavity can have several effects. It may serve as a means to protect QB from external disturbance, helping to preserve the delicate quantum state of QB. Actually, the mediated cavity can assist in maintaining coherence and reducing decoherence, which are crucial factors in quantum systems. So, this mediation can lead to more controlled and efficient energy transfer processes, potentially optimizing the charging performance of QB. Moreover, the mediated cavity helps to manage the interactions, allowing for better manipulation and utilization of quantum properties during the charging process.

Motivated by this, we propose a theoretical scheme for a QB consisting of an atom-cavity interacting system under a structured (bosonic) reservoir in the non-Markovian regime \cite{model1,model2,model3,model4}. We scrutinize a multi-parameter regime for the strength of cavity-reservoir coupling and demonstrate how these parameters affect the performance of the proposed QB. This scheme is simple yet complex enough for our goal, however, it is feasible for practical implementations.

\section{cavity-mediated charging performance of QB}\label{sec:2}
\begin{figure}[t]
\centering
    \includegraphics[width =0.60 \linewidth]{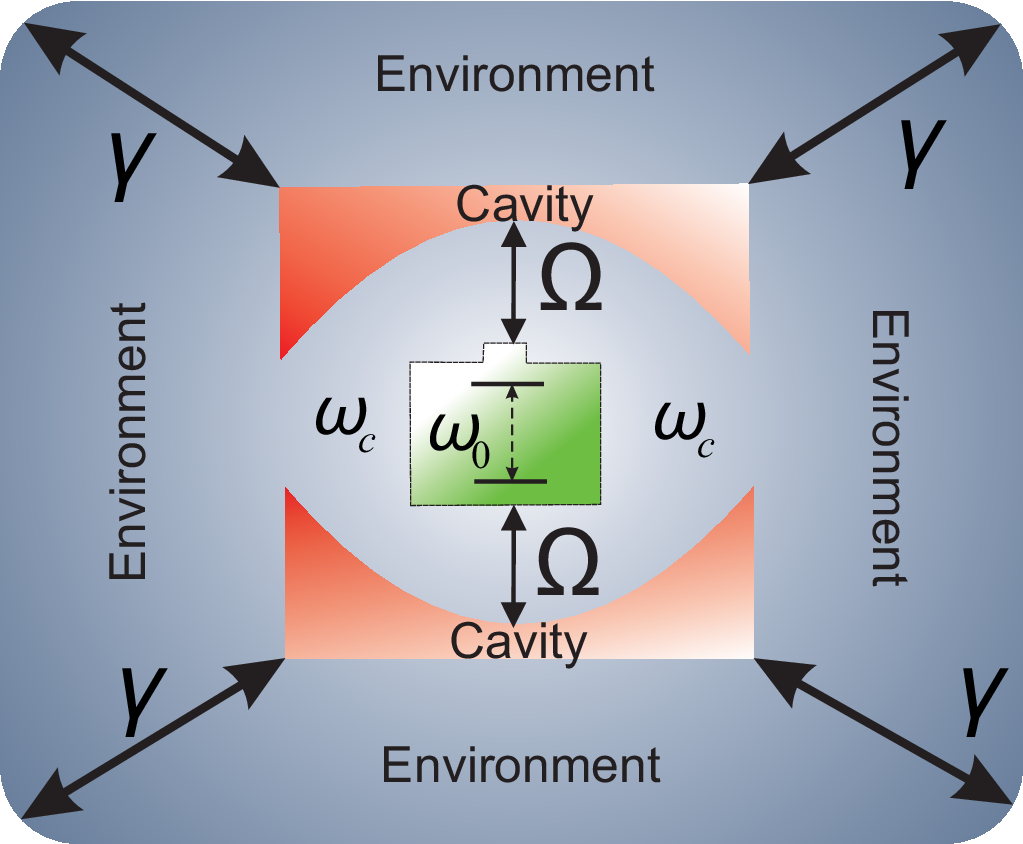}
    \caption{A visual representation of the considered set-up for cavity-mediated charging process of QB.}
    \label{Fig1}
  \end{figure}

The general system we are interested in studying includes a two-level system (qubit) considered a QB, a single-mode cavity, and a structured bosonic environment. A visual representation of our proposed scheme is illustrated in Fig. \ref{Fig1}.
The cavity acts as a mediator between the QB and the environment.  To find the conditions under which the QB can be charged with better efficiency \cite{qb11}, two different scenarios for the environment will be considered--one is an environment with memory and the other is a memory-less environment. The Hamiltonian of the composite system in the rotating wave approximation can be written as \cite{Maniscalco2008,Francica2009}
\begin{equation}\label{total hamiltonian}
H=H_0+f(t) H_I.
\end{equation}

In the above equation, the function $f(t)$ is the regulator of QB charging time. Its outcome equals $1$ for $t \in [0,\tau)$ with $\tau$ denoting the charging time of QB and is $0$ for other values of $t$. In other words, this function plays the role of an on-and-off switch for the interaction of subsystems. It is assumed that at time $t=0$, the QB will be connected to the cavity as well as the cavity to the environment. In the considered scenario, there exists no interaction between QB and the environment. So, energy transfer from the environment to QB during the charging period $[0, \tau )$ takes place through the cavity, which plays the role of a mediator. At the end of the charging process, i.e. at time $\tau$, QB and cavity are disconnected from cavity and environment respectively.

In Eq. \eqref{total hamiltonian}, $H_0$ is the free Hamiltonian of the total system, which can be written as
\begin{equation}\label{free hamiltonian}
H_0=\omega_0 \sigma_+ \sigma_- +\omega_c a^{\dagger} a+\sum_{k=0}^{\infty} \omega_k b_k^{\dagger} b_k,
\end{equation}
where $\sigma_+=\vert e \rangle \langle g \vert$ and $\sigma_-= \vert g \rangle \langle e \vert$ are the raising and lowering operators for qubit with $ \vert e \rangle$ and $\vert g \rangle$, which are excited and ground state of the qubit respectively. Besides, $\omega_0$ is the transition frequency of qubit. The second term in Eq. \eqref{free hamiltonian} represents the free Hamiltonian of the cavity with $a$  and  $a^{\dag}$ being the annihilation and creation operators of the cavity and $\omega_c$ is the transition frequency of the cavity. Herein, the resonance condition is considered $\omega_c=\omega_0$. The third term of the mentioned equation shows the free Hamiltonian of the environment in which $b_k$ and $b_k^{\dag}$ are the annihilation and creation operators for $k$th mode of the environment with frequency $\omega_k$. 

In the last term of Eq. \eqref{total hamiltonian}, $H_I$ denotes the interaction term of the Hamiltonian that describes the qubit-cavity and cavity-environment interaction given by
\begin{equation}
H_I=\Omega\left(\sigma^{+} a+\sigma^{-} a^{\dagger}\right)+\sum_{k=0}^{\infty} g_k\left(a b_k^{\dagger} + a^{\dagger} b_k \right),
\end{equation}
where $\Omega$ is the QB-cavity coupling strength and $g_k$ is the coupling strength between the cavity and $k$th mode of the environment.

Here, the time evolution of the total system will be calculated with a single excitation, given that the reservoir is in a vacuum state. So, the initial state of the total system can be considered as
\begin{equation}\label{instate}
\vert \psi(0) \rangle =\left[ c_1(0)\vert g_B \rangle \vert 1_c \rangle + c_2(0) \vert e_B \rangle \vert 0_c \rangle \right] \otimes \vert 0_k\rangle_{\mathcal{E} },
\end{equation}
where $\vert 0_c \rangle$ and $\vert 1_c \rangle$ are the vacuum and single-mode excitation states of the cavity respectively, and $\vert 0_k \rangle_\mathcal{E}$ represents the vacuum state of the bosonic environment. Moreover, $c_1(0)$ and $c_2(0)$ are the probability amplitudes.  So, the time-evolved state spanned by a single excitation basis state can be obtained as (see appendix \ref{Appendix A})
\begin{equation}\label{state}
\begin{aligned}
\vert \psi(t) \rangle =& [c_1(t) \vert g_B, 1_c \rangle + c_2(t)\vert e_B, 0_c \rangle]\otimes \vert 0_k \rangle_{\mathcal{E}} \\
& +\sum_k c_k(t)\vert g_B,0\rangle \otimes \vert 1_k \rangle_{\mathcal{E}},
\end{aligned}
\end{equation}
where $\vert 1_k \rangle_\mathcal{E}$ is the state of the bosonic environment with excitation in $k$th mode.

\section{Structured with-memory environment}
Let the reservoir be assumed to have a Lorentzian spectrum as $J(\omega)=\frac{\gamma}{2 \pi} \frac{\lambda^2}{(\omega_0-\omega)^2+\gamma^2}$, where $\gamma$ is the effective
coupling strength and $\lambda$ is the width of the spectrum which has the inverse relation with memory time of environment $\tau_\mathcal{E}$, i.e. $\lambda^{-1}=\tau_\mathcal{E}$. In this work, we study two different situations. One when the environment has memory $\tau_\mathcal{E} \neq 0$ and the other when it is memory-less $\tau_\mathcal{E}=0$.

Let us assume that the QB is initially empty. Hence, we set $c_1(0)=1$ and $c_2(0)=0$ in Eq. \eqref{instate}, implying that the initial state of the whole system is $\vert g_B, 1_c \rangle \otimes \vert 0_k \rangle_\mathcal{E} $. If the partial trace is taken over the cavity and environment, then the reduced density operator of the QB at time $\tau$ can be written as
\begin{equation}
\rho_B(\tau)=\vert c_2(\tau) \vert^{2} \vert e \rangle  \langle e \vert + \left[ 1-\vert c_2(\tau) \vert^{2} \right] \vert g \rangle \langle g \vert,
\end{equation}
where 
$c_2(\tau)=\kappa(\tau)c_2(0)$, with
\begin{equation}\label{ktav}
\kappa(\tau)=\mathcal{L}^{-1}\left(\frac{-i \Omega}{s^2+\Omega^2+\frac{\lambda \gamma s}{2(s+\lambda)}}\right),
\end{equation}
in which $\mathcal{L}^{-1}$ is the inverse Laplace transformation.

In order to study the charging process of the QB, we first analyze the dynamics in the context of Markovian (without memory) and non-Markovian (with memory) aspects in the presence of the environment with memory $\tau_{\mathcal{E}}\neq 0$. Markovian dynamics are those in which information continuously flows from a system to its environment during the evolution, while non-Markovian dynamics are those in which information backflow from the environment to the system \cite{nonmarkov1,nonmarkov2}. The Breuer--Laine--Piilo (BLP) measure is used here to quantify the degree of non-Markovianity of the quantum evolution \cite{nonmarkov2}.  Based on the trace distance between two distinct quantum states $\rho_1(t)$ and $\rho_2(t)$, namely $D(\rho_1(t), \rho_2(t))=\frac{1}{2} \textmd{tr} \vert \rho_1(t) - \rho_2(t) \vert$, the BLP measure quantifies the amount of non-Markovianity $\mathcal{N}$ as
\begin{equation}\label{non}
\mathcal{N}=\max _{\rho_1(0), \rho_2(0)} \int_{\sigma>0} dt~\sigma\left(t, \rho_1(0), \rho_2(0)\right),
\end{equation}
where $\sigma(t,\rho_1(0),\rho_2(0))=\frac{d}{dt}D(\rho_1(t),\rho_2(t))$ is the rate of change of the trace distance. 

Using a large sample set of pairs of initial states and strong numerical evidence, Ref. \cite{nonmarkov3} determines that Eq. \eqref{non} has a maximum value for states $\rho_1(0)=\vert e \rangle \langle e \vert$ and $\rho_2(0)=\vert g \rangle \langle g \vert$. Based on these two initial states, we obtain $\sigma\left(t, \rho_1(0), \rho_2(0)\right)=\frac{d}{dt} \kappa(t)$. If $\sigma\left(t, \rho_1(0), \rho_2(0)\right)>0$, then we have $\mathcal{N}\neq 0$.

\begin{figure}[t]
\centering
    \includegraphics[width =0.9 \linewidth]{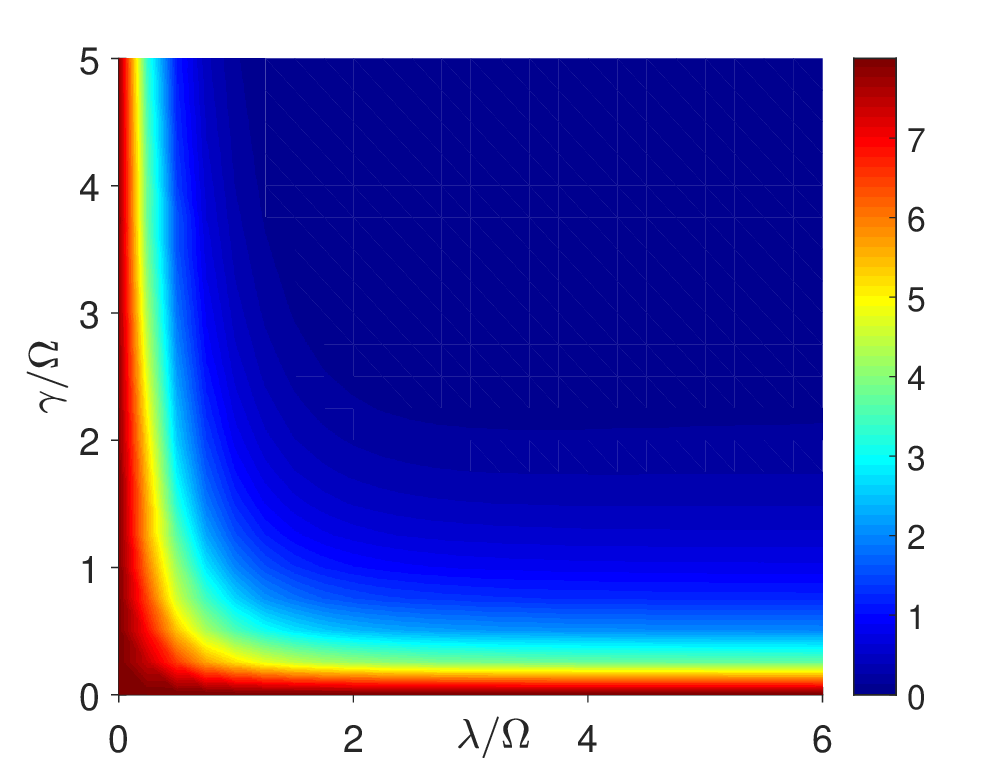}
    \vspace*{-5mm}
    \caption{Non-Markovianity $\mathcal{N}$ as functions of $\gamma/\Omega$ and $\lambda/\Omega$. }
    \label{Fig2}
  \end{figure}

In Fig. \ref{Fig2}, the non-Markovianity is plotted as functions of $\gamma/\Omega$ and $\lambda/\Omega$. From this figure, it is clear that these two parameters have significant effects on the evolution from the Markovian and non-Markovian aspects, one is the ratio of the width of the spectrum $\lambda$ to the QB-cavity coupling strength $\Omega$, and the other is the ratio of cavity-environment coupling strength $\gamma$ to QB-cavity coupling strength $\Omega$. On one hand, we observe that the strong coupling between the QB and the cavity $\Omega$ leads to a greater non-Markovian feature of the evolution. On the other hand, from the dependence of the width of the spectrum $\lambda$  to the memory time $\tau_\mathcal{E}$ of the structured environment as $\lambda^{-1}=\tau_{\mathcal{E}}$, we see that the longer memory time of the environment increases the amount of non-Markovianity. We also find that with the increase in the cavity-environment coupling strength $\gamma$, the amount of non-Markovianity decreases.  It has been shown that QBs in the non-Markovian regime have the best efficiency in the charging process \cite{decoh4}, which is why we have first studied the evolution from both Markovian and non-Markovian perspectives.

Let us return to the main issue, which is the cavity-mediated charging performance of QBs.  It is straightforward to calculate the amount
of energy that the battery obtains during the charging process as $\Delta E_B (\tau)=E_B(\tau)-E_B(0)$, where $E_B(\tau)=\textmd{tr}[\rho_B(\tau)H_B]$. Since the battery is initially empty, we have $\Delta E_B(\tau)=\omega_0 \vert c_2(\tau) \vert^2$.

\begin{figure}[t]
\begin{minipage}[t]{1\linewidth}
    \centering
    \includegraphics[width=0.9\textwidth]{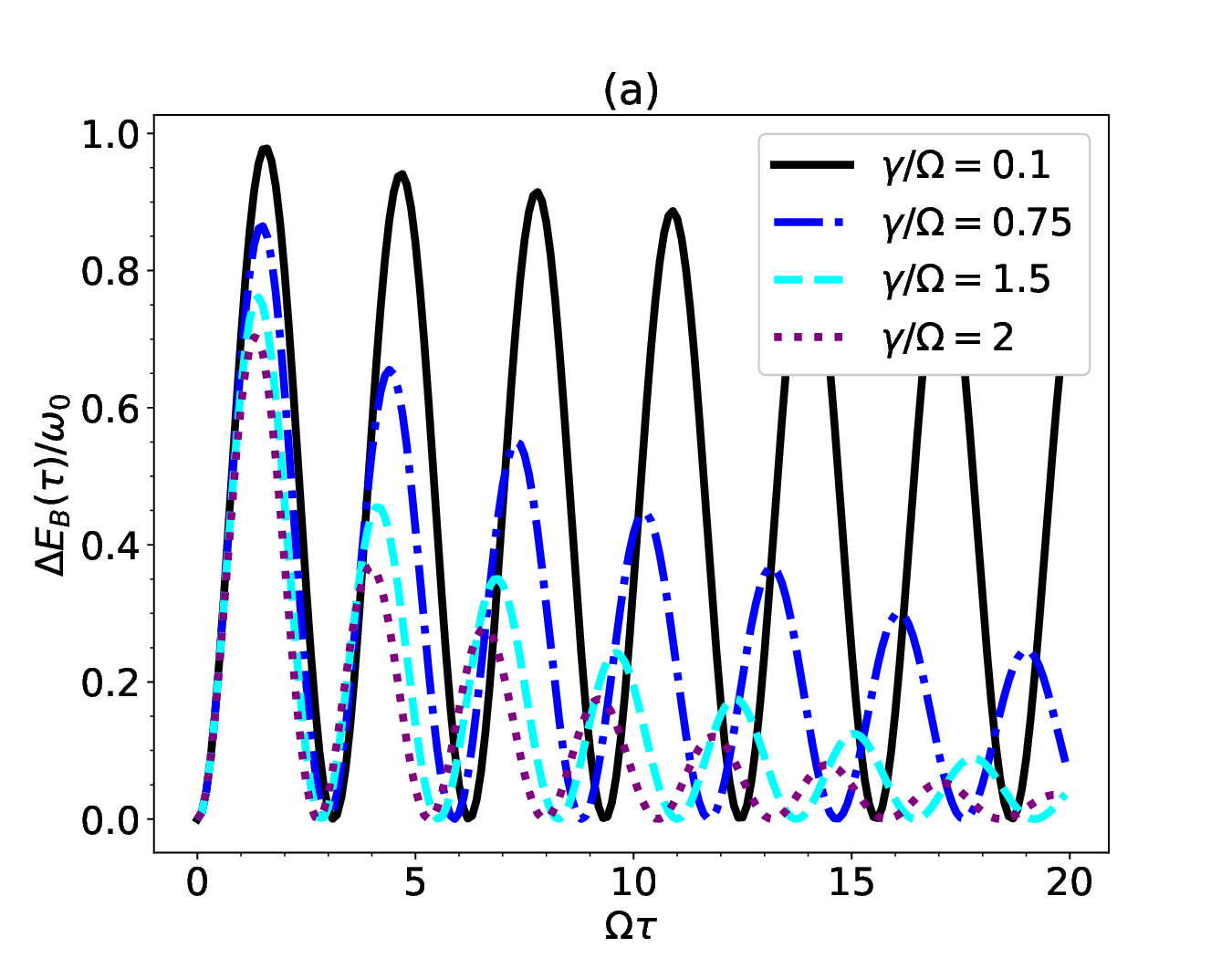}
\end{minipage}
\vspace*{-5mm}
\begin{minipage}[t]{1\linewidth}
    \centering
    \includegraphics[width=0.9\textwidth]{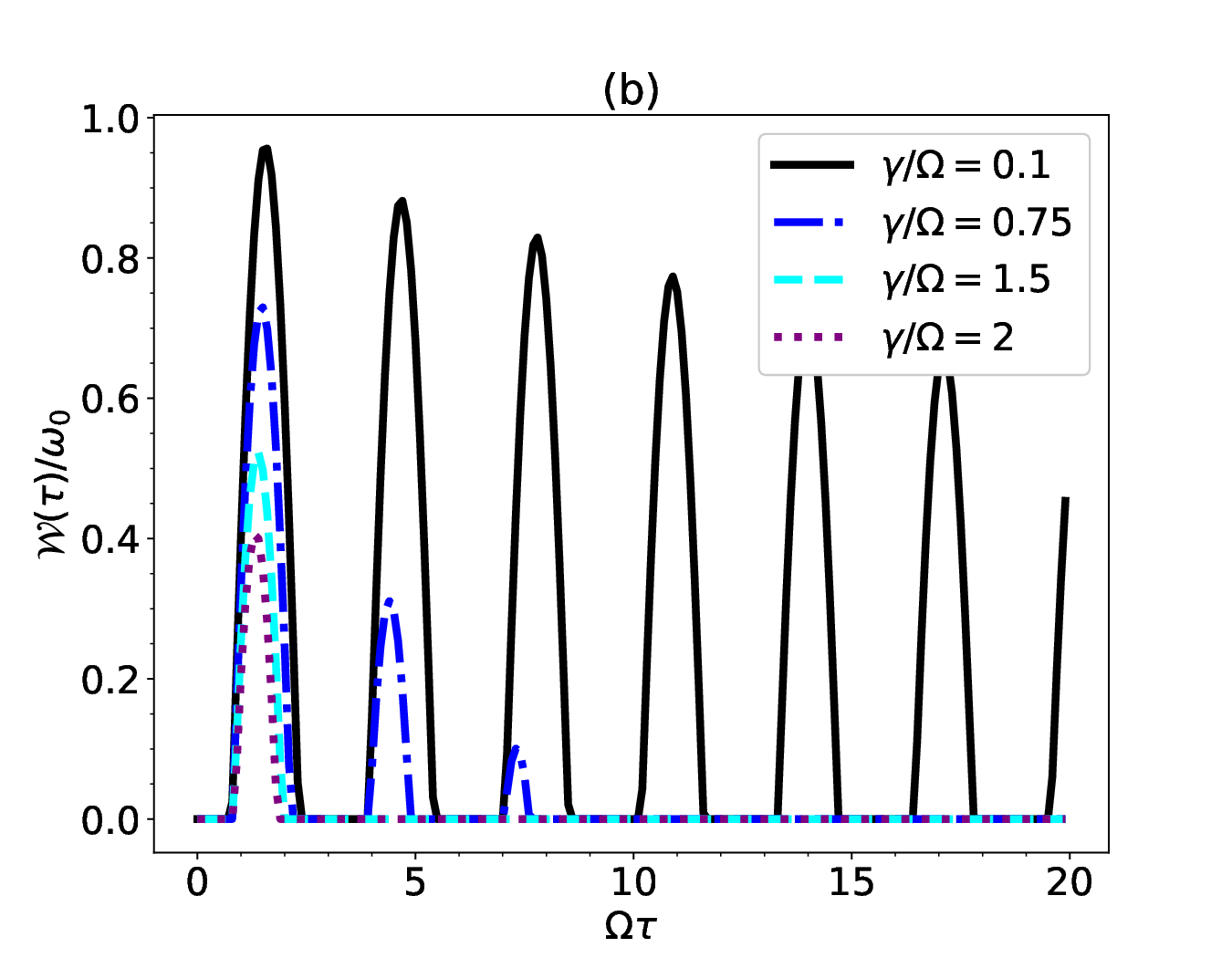}
\end{minipage}
\vspace*{-3mm}
\caption{(\textbf{a}) Dynamics of stored energy in QB during charging process  $\Delta E_B(\tau)$ and (\textbf{b}) ergotropy $\mathcal{W}(\tau)$ as a function of dimensionless parameter $\Omega \tau$ in non-Markovian regime with $\lambda/\Omega =0.1$ for different values of $\gamma / \Omega$.}
\label{Fig3}
\end{figure}

In Fig. \ref{Fig3}(a), we have plotted the non-Markovian dynamics of $\Delta E_B(\tau)$ in terms of dimensionless parameter $\Omega \tau$ with $\lambda=0.1 \Omega$ in the presence of structured environment with memory. It is observed that, during the charging process, the energy obtained by the QB increases with the enhancement in the QB-cavity coupling strength $\Omega$, while it decreases with the strengthening of the cavity-environment coupling $\gamma$.

The ergotropy describes the maximum amount of energy that can be extracted from the QB under cyclic unitary operation after it has been charged for a certain period of time $\tau$ \cite{Allahverdyan2004}, defined by (see appendix \ref{Appendix B}) 
\begin{equation}\label{Ergo}
\mathcal{W}=\textmd{tr}(\rho_B(\tau)H_B)-\textmd{tr}(\sigma_{\rho_B}H_B),
\end{equation}
where $\sigma_{\rho_B}$ is called the passive state of $\rho_B$ and it is a state from which no work can be extracted under cyclic unitary operations. From Eq. \eqref{Ergo}, the ergotropy takes the following form for the cavity-mediated scenario
\begin{equation}
\mathcal{W}(\tau)=\omega_0(2 \vert c_2(\tau)\vert^2 -1)~\Theta (\vert c_2(\tau)\vert^2-\frac{1}{2}),
\end{equation}
where $\Theta(x-x_0)$ indicates the Heaviside function. 

The time-evolution of ergotropy is plotted in Fig. \ref{Fig3}(b) as a function of $\Omega \tau$ in a non-Markovian regime with memory $\lambda = 0.1 \Omega$. We find that the most energy can be extracted from the QB  when the QB-cavity coupling strength $\Omega$ is greater than the cavity-environment coupling strength $\gamma$.

The maximum stored energy in QB $\Delta E_B^{max}$ and the maximum ergotropy $\mathcal{W}_{max}$ can then be used to evaluate the charging performance of the cavity-mediated scenario of QBs. They are given by
\begin{equation}
\Delta E_B^{max}=\max_\tau [\Delta E_B (\tau)], \quad \mathcal{W}_{max}=\max_\tau [\mathcal{W} (\tau)],
\end{equation}
where the optimization is done over the charging time $\tau$.  Note that greater $\Delta E_B^{max}$ and $\mathcal{W}_{max}$ are required for optimal charging of the QB.

\begin{figure}[t]
\begin{minipage}[t]{1\linewidth}
    \centering
    \includegraphics[width=0.9\textwidth]{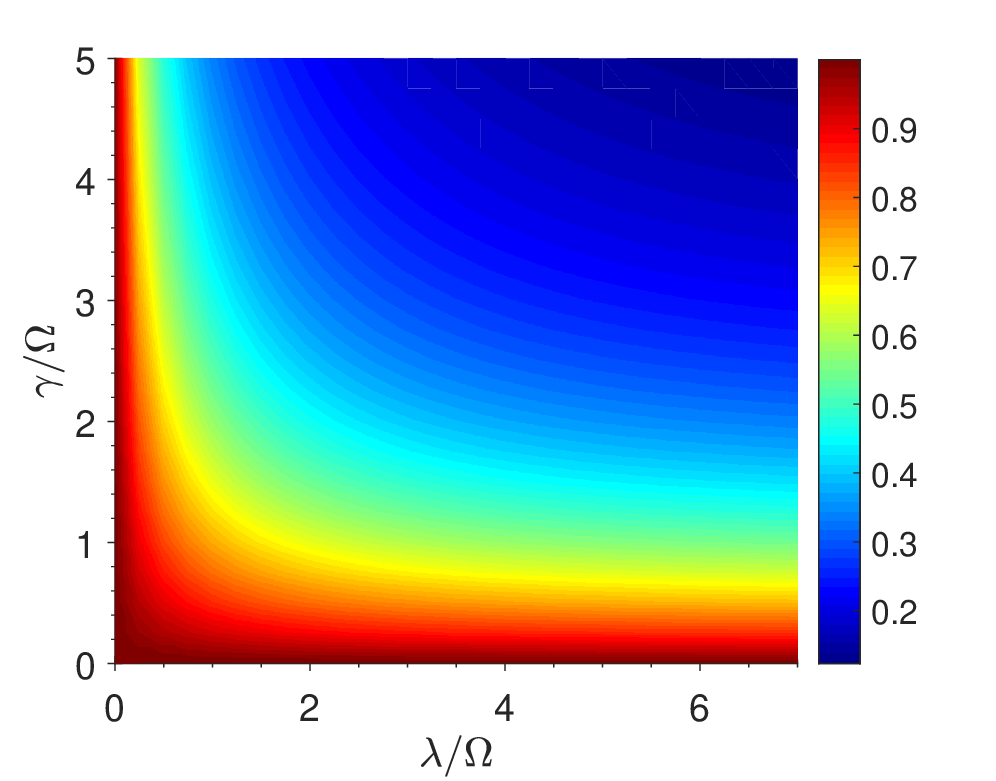}
\end{minipage}
\vspace*{-5mm}
\begin{minipage}[t]{1\linewidth}
    \centering
    \includegraphics[width=0.9\textwidth]{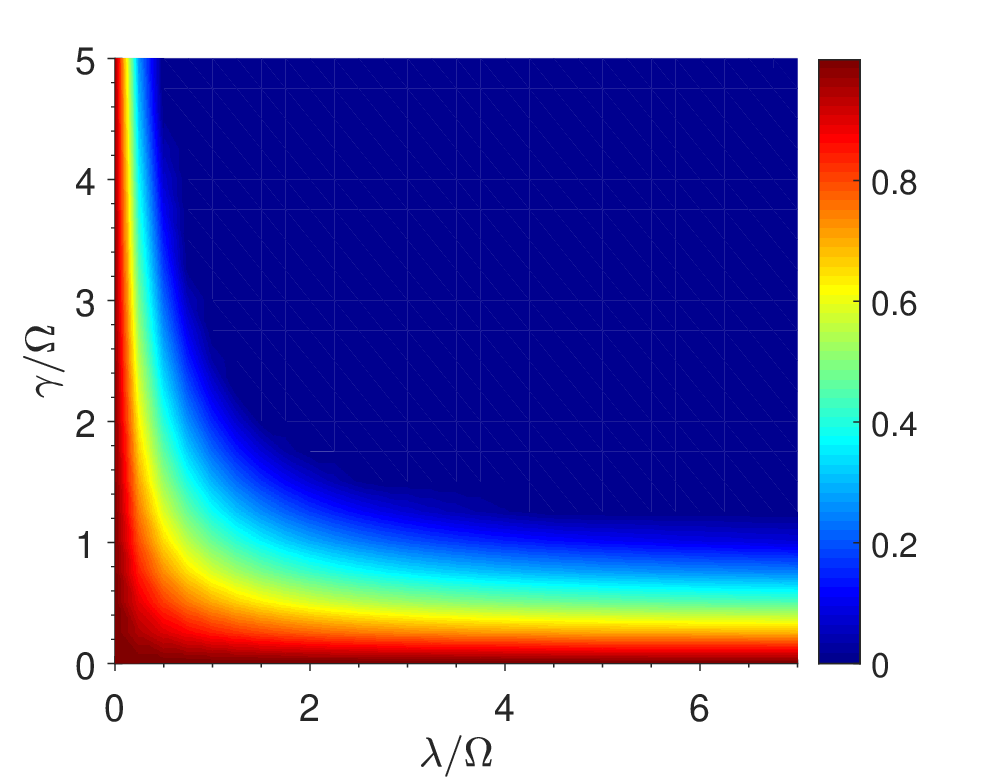}
\end{minipage}
\vspace*{-3mm}
\caption{(\textbf{a}) Maximum stored energy in QB, $\Delta E_B^{max}$, and (\textbf{b}) Maximum ergotropy, $\mathcal{W}_{max}$, as functions of $\gamma/\Omega$ and $\lambda/\Omega$.}
\label{Fig4}
\end{figure}

In Fig. \ref{Fig4}(a), the maximum energy stored in QB $\Delta E_B^{max}$ is plotted in terms of $\gamma/\Omega$ and $\lambda/\Omega$. As can be seen, the maximum stored energy in QB reaches its highest value for stronger QB-cavity coupling $\Omega$. It can also be said that for larger memory time $\tau_{\mathcal{E}}$ of the structured environment (small value of $\lambda$), $\Delta E_B^{max}$ has the highest value. Moreover, Fig. \ref{Fig4}(b) illustrates $\mathcal{W}_{max}$ as  functions of $\gamma/\Omega$ and $\lambda/\Omega$. We find that the highest value of $\mathcal{W}_{max}$ is obtained for smaller values of $\lambda$ (longer memory time of environment $\tau_{\mathcal{E}}$). It is also observed that $\mathcal{W}_{max}$ increases by amplifying the QB-cavity coupling $\Omega$ and decreases with increasing the effective cavity-environment coupling $\gamma$.

\section{Structured memory-less environment }
Let us consider the situation in which the memory time of the environment is zero. For the environment to be memory-less ($\tau_\mathcal{E}=0$), it is necessary to set $\lambda \rightarrow \infty$. So, in this section, we follow all the steps that we derived in the previous sections, but we just consider the limit $\lambda \rightarrow \infty$ in all of the calculations. Hence, we can rewrite Eq. \eqref{ktav} as
\begin{equation}
\kappa(\tau)=- 4 i \exp[-\frac{\mathcal{R}}{4}\tau]\sinh[\frac{\mathcal{R}}{4}\tau],
\end{equation}
where $\mathcal{R}=\sqrt{\gamma^2-16 \Omega^2}$. From the BLP measure for non-Markovianity, it is observed that the evolution is non-Markovian for $\gamma / \Omega <4$. Since the charging performance of QBs is more optimal in the non-Markovian regime, therefore, we will study here the charging performance in the non-Markovian regime.

\begin{figure}[t]
\begin{minipage}[t]{1\linewidth}
    \centering
    \includegraphics[width=0.9\textwidth]{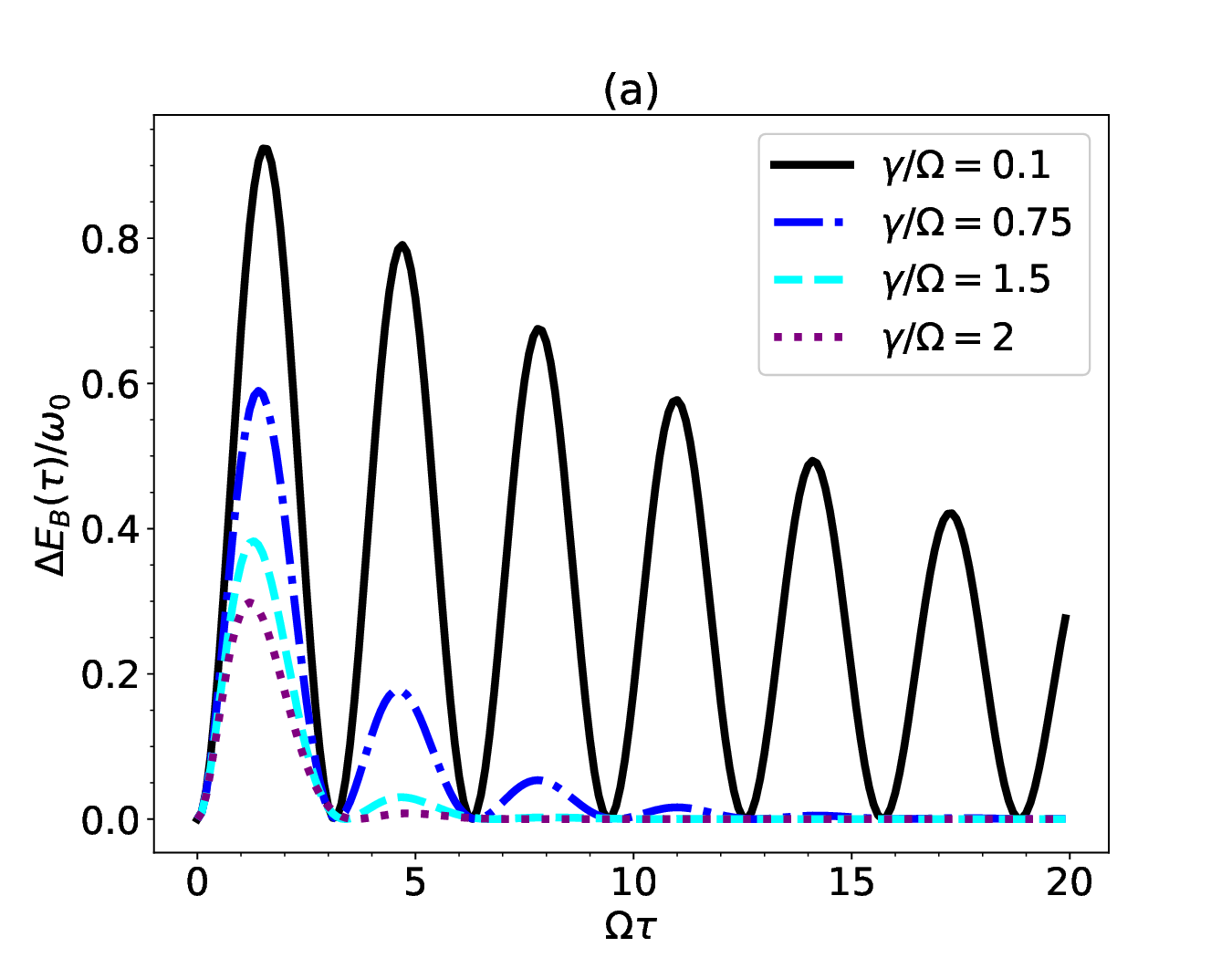}
\end{minipage}
\vspace*{-5mm}
\begin{minipage}[t]{1\linewidth}
    \centering
    \includegraphics[width=0.9\textwidth]{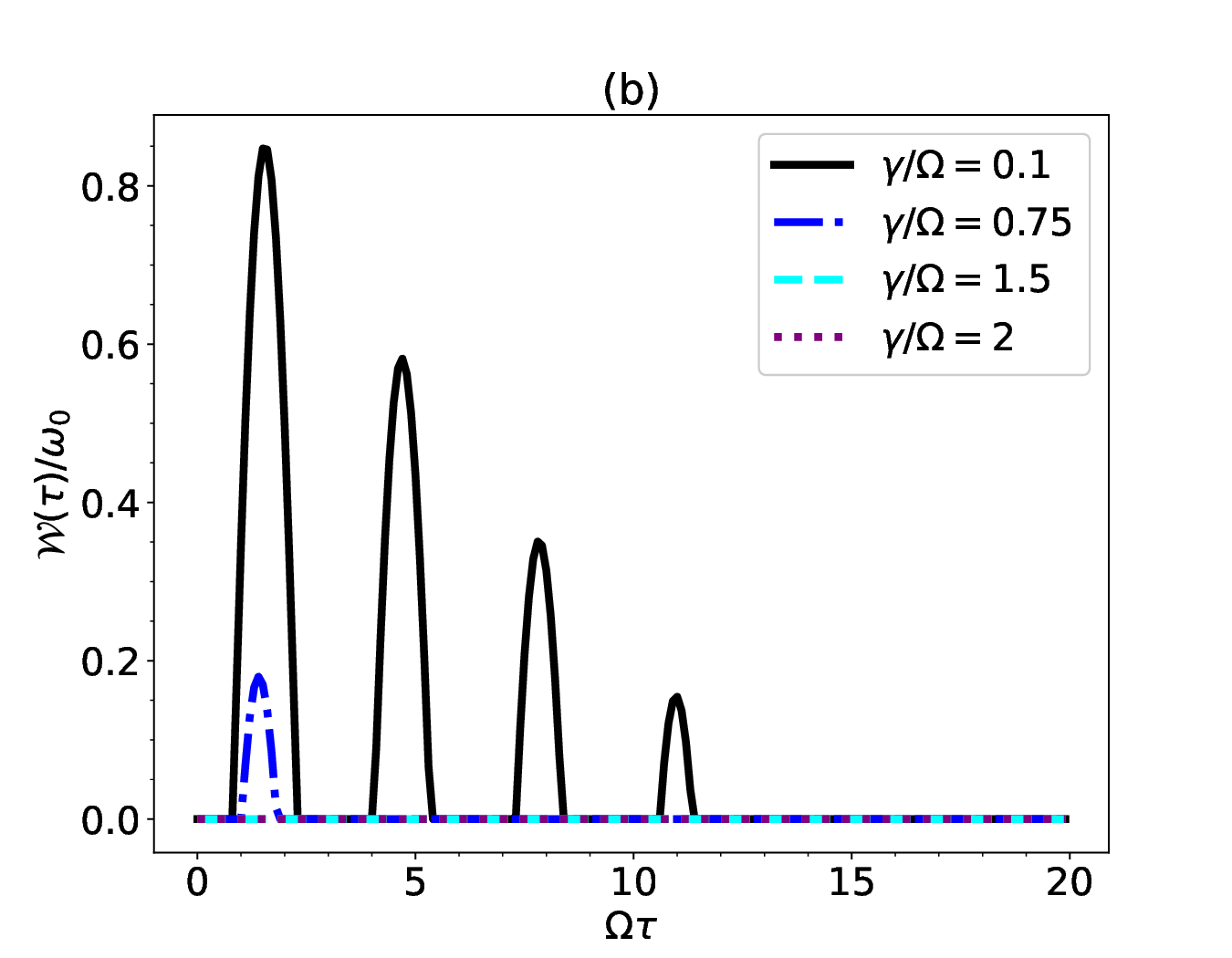}
\end{minipage}
\vspace*{-3mm}
\caption{(\textbf{a}) Dynamics of stored energy in QB during charging process  $\Delta E_B(\tau)$ and (\textbf{b}) ergotropy $\mathcal{W}(\tau)$  as a function of $\Omega \tau$ in non-Markovian regime with memory-less environment $\lambda \rightarrow \infty$ for different values of $\gamma / \Omega$.}
\label{Fig5}
\end{figure}

\begin{figure}[t]
\begin{minipage}[t]{1\linewidth}
    \centering
    \includegraphics[width=0.9\textwidth]{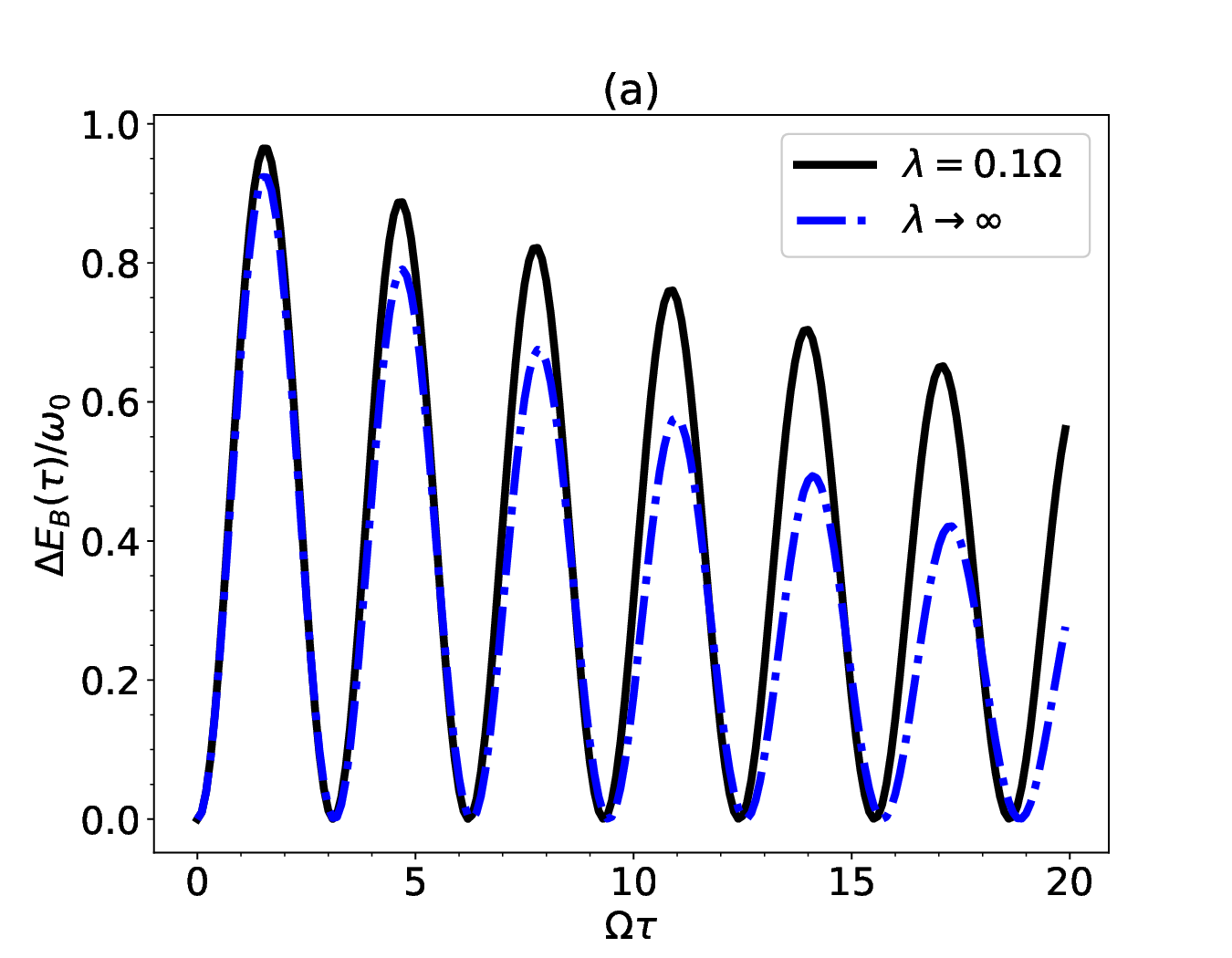}
\end{minipage}
\vspace*{-5mm}
\begin{minipage}[t]{1\linewidth}
    \centering
    \includegraphics[width=0.9\textwidth]{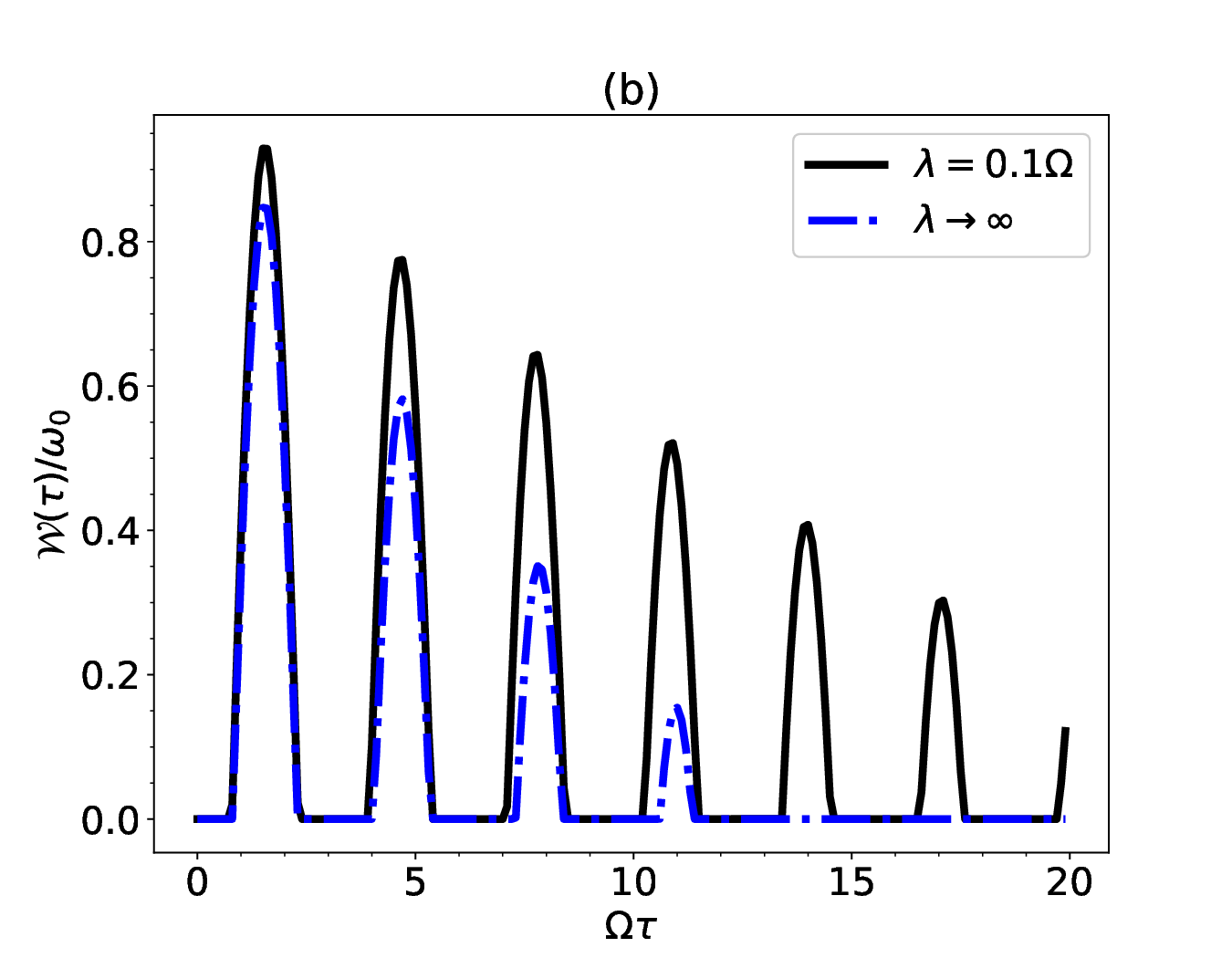}
\end{minipage}
\vspace*{-3mm}
\caption{(\textbf{a}) Dynamics of stored energy in QB during charging process  $\Delta E_B(\tau)$ and (\textbf{b}) ergotropy $\mathcal{W}(\tau)$ as a function of $\Omega \tau$ in non-Markovian regime for different types of environments with $\gamma=0.1 \Omega$.}
\label{Fig6}
\end{figure}

\begin{figure}[t]
\begin{minipage}[t]{1\linewidth}
    \centering
    \includegraphics[width=0.9\textwidth]{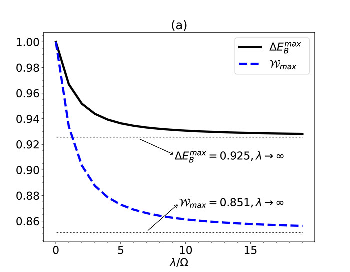}
\end{minipage}
\vspace*{-5mm}
\begin{minipage}[t]{1\linewidth}
    \centering
    \includegraphics[width=0.9\textwidth]{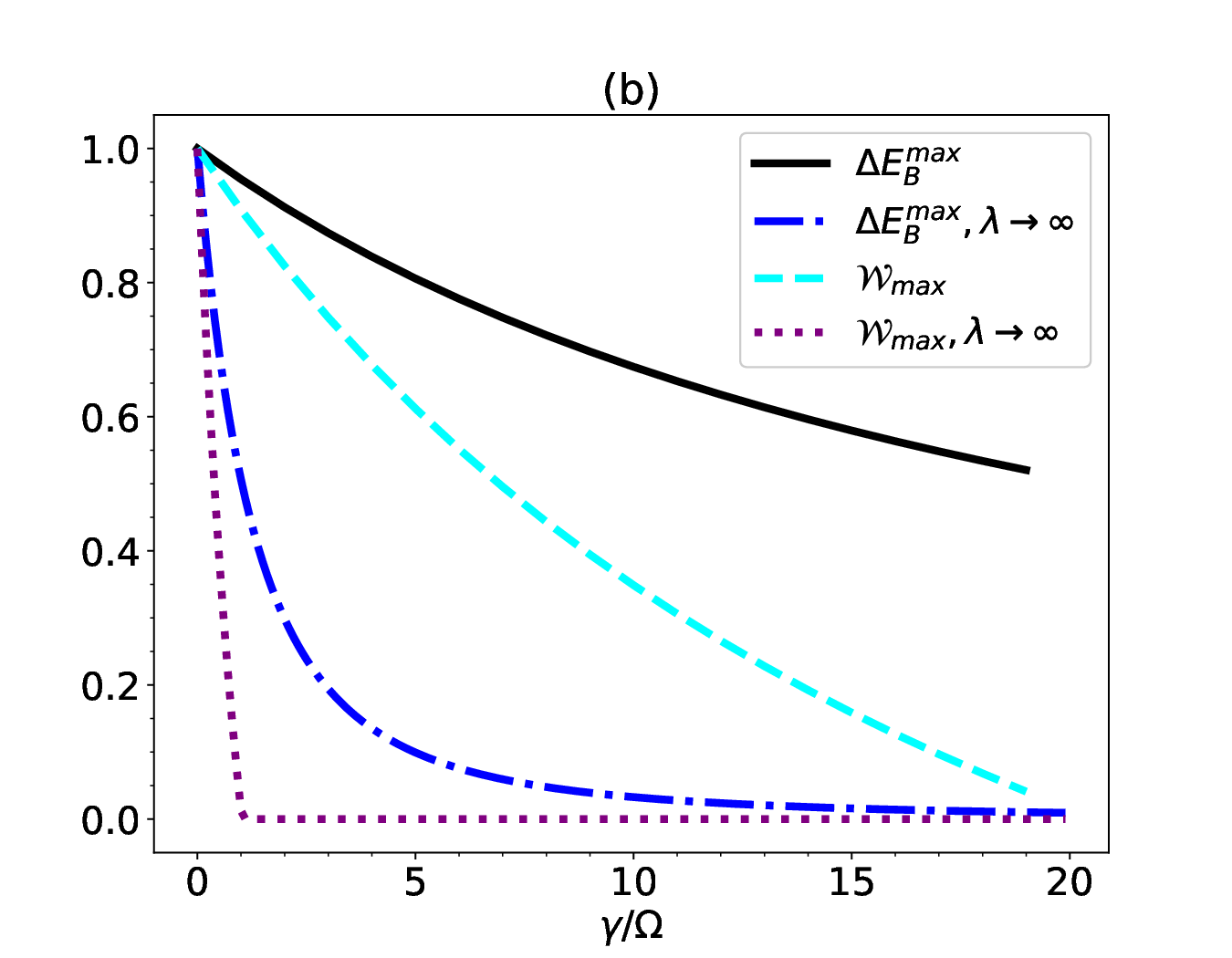}
\end{minipage}
\vspace*{-3mm}
\caption{ Comparative plots of maximum stored energy $\Delta E_B^{max}$ and maximum ergotropy $\mathcal{W}_{max}$ from the point of view of memory-less ($\lambda \rightarrow \infty$) and with-memory  environments as a function of (\textbf{a}) $\lambda/ \Omega$ (with $\gamma=0.1 \Omega$) and (\textbf{b}) $\gamma/ \Omega$ (with $\lambda=0.1\Omega$) in non-Markovian regime.}
\label{Fig7}
\end{figure}

In Fig. \ref{Fig5}(a), the stored energy in QB  is plotted as a function of dimensionless parameter $\Omega \tau$ in the memory-less environment for different values of $\gamma /\Omega$. As can be seen, $\Delta E_B (\tau)$ decreases by increasing effective coupling between cavity and environment $\gamma$. It can also be noticed that the stored energy increases with raising the QB-cavity coupling strength $\Omega$. Meanwhile, the ergotropy $\mathcal{W}$ is plotted in terms of $\Omega \tau$ for different values of $\gamma /\Omega$ in Fig. \ref{Fig5}(b). The ergotropy decreases with increasing $\gamma$ and finally tends to zero for larger values of $\gamma$. Notice that the ergotropy achieves its highest value for greater values of QB-cavity coupling strength $\Omega$.

\section{Comparison between with-memory and memory-less environments}
Here, we examine which of the environments with memory or memory-less is more efficient in the charging process of QB. To compare the stored energy and ergotropy in these two different environments, we consider the same effective coupling $\gamma$ between the environment and the cavity. 
\\
\\
Fig. \ref{Fig6}(a) shows the evolution of stored energy $\Delta E_B(\tau)$ in QB for different types of environments for the same cavity-environment effective coupling $\gamma=0.1 \Omega$. Since the stored energy $\Delta E_B(\tau)$ in the presence of the memory-less environment (blue dot-dashed line) is less than the stored energy in the presence of the with-memory environment (black solid line), the optimal charging performance of QB happens for the with-memory environment. In Fig. \ref{Fig6}(b), the ergotropy $\mathcal{W}(\tau)$ is plotted as a function of $\Omega \tau$ for different types of environments with $\gamma=0.1 \Omega$. In agreement with the result of Fig. \ref{Fig6}(a), the ergotropy in the with-memory environment is greater than in the memory-less environment.

To make it clear that the charging performance is more optimal in a with-memory environment, we plot $\Delta E_B^{max}$ and $\mathcal{W}_{max}$ in terms of $\lambda / \Omega$ [see Fig. \ref{Fig7}(a)] and $\gamma/\Omega$ [see Fig. \ref{Fig7}(b)]. From Fig. \ref{Fig7}(a), it is observed that for the memory-less environment $\lambda \rightarrow \infty$ with $\gamma = 0.1 \Omega$, the value of maximum stored energy is equal to $0.925$ (the upper thin dotted line in this plot indicates this value).  A comparison of this value with the maximum stored energy in the QB  reveals that the value of $\Delta E_B^{max}$ in the with-memory environment (black solid line) is higher than that in the memory-less environment. Moreover, one can see that the maximum value of ergotropy for the memory-less environment $\lambda \rightarrow \infty$  is equal to $0.851$ (the lower thin dotted line in this plot indicates this value). Also, we find that the maximum value of the ergotropy $\mathcal{W}_{max}$ in the presence of an environment with memory  (blue dashed line) is greater than that in a memory-less environment.

In Fig. \ref{Fig7}(b), we sketch $\Delta E_B^{max}$ and $\mathcal{W}_{max}$ versus $\gamma/\Omega$.  From this figure, we notice that the maximum values of the stored energy and ergotropy for the with-memory environment ($\lambda=0.1\Omega$) are always greater than those in the memory-less environment $\lambda \rightarrow \infty$.

\section{Conclusion and outlook}
In this work, we have studied the cavity-mediated charging performance of QB in the presence of a structured bosonic environment. This study considered a practical model in which a cavity acts as a mediator between the QB and the structured environment. In other words, there exists no direct interaction between QB and structured bosonic environment. It was observed that, with increasing the coupling strength between the cavity and environment $\gamma$, the QB will not charge optimally and the amount of stored energy in the QB will decrease.  Furthermore, increasing the coupling strength between the QB and the cavity $\Omega$ optimizes the QB charging process and increases the amount of work that can be extracted from the QB.

We have also studied the effects of two different types of environments from the memory point of view on the charging process of QB individually. The environment has memory when its correlation time has a non-zero value, while the environment is memory-less when its correlation time is equal to zero. We found that for the with-memory environment, the stored energy and extractable work are greater than those in which the cavity interacts with a memory-less environment. Therefore, it can be concluded that it is possible to extract work from QB in a more efficient manner in with-memory environments.

\appendix

\section{Analytical solution for the system dynamics}
\label{Appendix A}
Here, we examine the model used for the operational investigation of QBs in more detail. The interaction Hamiltonian in the interaction picture can be written as  
\begin{equation}
\tilde{H}_I=\Omega(\sigma_+ a + \sigma_- a^{\dag}) + \sum_{k=0}^{\infty} g_k (a b_k^{\dag}e^{i \delta_k t}+a^{\dag} b_k e^{-i \delta_k t}).
\end{equation}
By considering the above Hamiltonian and inserting Eq. \eqref{state} into the Schr\"{o}dinger equation, the following coupled differential equations are derived as
\begin{equation}
\begin{aligned}
\frac{d}{d t} c_1(t) & =-i \Omega c_2(t)-i \sum_k g_k e^{-i \delta_k t} c_k(t),\\
\frac{d}{d t} c_2(t) & =-i \Omega c_1(t), \\
\frac{d}{d t} c_k(t) & =-i g_k e^{i \delta_k t} c_1(t),
\end{aligned}
\end{equation} 
where the resonance condition is considered $\omega_c=\omega_0$ and $\delta_k=\omega_k-\omega_0$. Considering the correlation function $f(t,t^{\prime})=\sum_k \vert g_k \vert^{2} e^{- i \delta_k (t-t^{\prime})}$, the above coupled differential equations can be solved using the Laplace transformation method as 
\begin{equation}
\begin{aligned}
 c_1(t) & =\mathcal{L}^{-1}\left(\frac{c_1(0)-i \frac{\Omega}{s}c_2(0)}{s+\frac{\Omega^{2}}{s} + \frac{\lambda \gamma}{2(s+\lambda)}}\right)
\end{aligned}
\end{equation} 
and
\begin{equation}
\begin{aligned} 
c_2(t) & =\mathcal{L}^{-1}\left(\frac{c_2(0)}{s}-i \frac{\Omega}{s}\frac{c_1(0)-i \frac{\Omega}{s}c_2(0)}{s+\frac{\Omega^{2}}{s} + \frac{\lambda \gamma}{2(s+\lambda)}}\right).
\end{aligned}
\end{equation} 

Note that the probability amplitude $c_k(t)$ can be obtained through the normalization condition.

\section{The ergotropy}
\label{Appendix B}
The QB as a quantum system can be described by the density matrix $\rho_B$ and the Hamiltonian $H_B$, which their spectral decompositions are given by
\begin{equation}
 \rho_B =\sum_n r_n \vert r_n \rangle \langle r_n \vert, \quad r_1 \geq r_2 \geq... \geq r_n 
\end{equation}
and
\begin{equation}
H_B =\sum_n \varepsilon_n \vert \varepsilon_n \rangle \langle \varepsilon_n \vert, \quad \varepsilon_1 \leq \varepsilon_2 \leq ... \leq \varepsilon_n,
\end{equation} 
where the sets $\lbrace \vert r_n \rangle \rbrace$ and  $\lbrace \vert \varepsilon_n \rangle\rbrace$ are eigenvectors of $\rho_B$ and $H_B$ respectively. Moreover, $r_n$'s and $\varepsilon_n$'s are eigenvalues of $\rho_B$ and $H_B$ respectively. The passive state $\sigma_B$ of $\rho_B$ is the state that can not extract work from it and is given by
\begin{equation}
\sigma_B=\sum_n r_n \vert \varepsilon_n \rangle \langle \varepsilon_n \vert.
\end{equation} 
Thereby, the ergotropy can be formulated as 
\begin{equation}
\mathcal{W}=\sum_{n,m} r_n \varepsilon_m \left( \vert \langle r_n \vert \varepsilon_m \rangle \vert^2 - \delta_{m,n} \right),
\end{equation}
where $\delta_{m,n}$ is the Kronecker delta function.

\section*{Acknowledgments}
D. Wang was supported by the National Natural Science Foundation of China (Grant No. 12075001), and Anhui Provincial Key Research and Development Plan (Grant No. 2022b13020004). S. Haddadi was supported by Semnan University under Contract No. 21270.

\section*{Disclosures}
The authors declare that they have no known competing financial interests.

\section*{Data availability}
No datasets were generated or analyzed during the current study.



\end{document}